\documentclass[conference]{IEEEtran}
\usepackage{siunitx}

\usepackage{cite}

\ifCLASSINFOpdf
   \usepackage[pdftex]{graphicx}
\else
   \usepackage[dvips]{graphicx}
\fi
\usepackage{amsmath, amssymb}
\usepackage{bm}

 \usepackage[caption=false,font=footnotesize]{subfig}
\begin{document}
%
\title{Visual Attention for Musical Instrument Recognition}

\author{\IEEEauthorblockN{
Karn Watcharasupat\IEEEauthorrefmark{1}\IEEEauthorrefmark{2}, 
Siddharth Gururani\IEEEauthorrefmark{1}
and Alexander Lerch\IEEEauthorrefmark{1}}
\IEEEauthorblockA{
\IEEEauthorrefmark{1}
Center for Music Technology\\
Georgia Institute of Technology\\
Atlanta, Georgia, USA 30318}
\IEEEauthorblockA{
\IEEEauthorrefmark{2}
School of Electrical and Electronic Engineering\\
Nanyang Technological University\\
Singapore 639798\\
Emails: karn001@e.ntu.edu.sg,  \{siddgururani, alexander.lerch\}@gatech.edu}
}





\maketitle

\begin{abstract}
In the field of music information retrieval, the task of simultaneously identifying the presence or absence of multiple musical instruments in a polyphonic recording remains a hard problem. Previous works have seen some success in improving instrument classification by applying temporal attention in a multi-instance multi-label setting, while another series of work has also suggested the role of pitch and timbre in improving instrument recognition performance. In this project, we further explore the use of attention mechanism in a timbral-temporal sense, à la visual attention, to improve the performance of musical instrument recognition using weakly-labeled data. Two approaches to this task have been explored. The first approach applies attention mechanism to the sliding-window paradigm, where a prediction based on each timbral-temporal `instance' is given an attention weight, before aggregation to produce the final prediction. The second approach is based on a recurrent model of visual attention where the network only attends to parts of the spectrogram and decide where to attend to next, given a limited number of `glimpses'.
\end{abstract}

\section{Introduction}
In musical pieces, musical instruments play a key role in the production of the music itself, and in defining the genre, mood, timbre, and many other aural characteristics of the piece. The ability to automate the identification of the presence or absence of instruments in a given recording can be helpful in several music information retrieval (MIR) fields such as music audio tagging, source separation, and automatic transcription. 

Such a task, known as musical instrument recognition, remains difficult for most musical pieces, which are often polyphonic and multi-instrumental. In particular, the challenges faced are\cite{DBLP:conf/ismir/GururaniSL19}: (a) the superposition in both timbral and temporal sense of the instrument sounds, (b) the wide variety of timbre produced by a family of instruments, and (c) the lack of well-annotated dataset for supervised learning. While the first two are inherent to the nature of the music and musical instruments, the last, however, continues to be tackled by the research community over the years. 

Recently, an open, moderately-sized dataset for multiple instrument recognition, OpenMIC-2018, was released \cite{eric_humphrey_2018_1492445}. The dataset provides \num{20000} audio clips annotated for the presence of 20 distinct instrument categories, along with 128-dimensional features extracted using VGGish \cite{VGGish}. The OpenMIC-2018 dataset, however, is weakly-labeled and has many missing labels across the instrument categories. It also suffers from significant positive-negative example imbalances in some classes, which complicates the training of networks using this dataset. Regardless, OpenMIC-2018 is currently the only dataset for musical instrument recognition which is moderately sized, multi-labeled, diverse, and polyphonic.

In this project, we explore the role of visual attention in improving musical instrument recognition for multi-instrument polyphonic music, through both deterministic attention similar to \cite{DBLP:conf/ismir/GururaniSL19} and stochastic attention similar to \cite{mnih2014recurrent}. The following section discusses related work in musical instrument recognition, music audio tagging, and neural attention. Section \ref{method} discusses the proposed models, followed by their evaluation in Section \ref{eval}. We discuss the results in Section \ref{results} and conclude the paper in Section \ref{conclusion}.

\section{Related Work}

\subsection{Musical Instrument Recognition}
Musical instrument recognition is a broad umbrella of tasks focused on identifying one or more instruments from a recording, which may be a solo performance, multi-instrument performance, or simply a brief note, etc. 

In recent years, the research has shifted from traditional feature analysis (see \cite{deng2008study}) towards the use of deep learning. Early deep learning works include \cite{han2016deep} which focused on multi-label instrument recognition in polyphonic music and \cite{DBLP:conf/ismir/LostanlenC16} which focused on single-instrument classification. Both of these works rely on convolutional neural networks (CNN) and found significant improvement compared to feature-based classification. Recent works that follow continue to rely to CNN, e.g.  \cite{juan_s_gomez_2018_1492481, takumi_takahashi_2018_1492477, DBLP:conf/ismir/TaenzerAMWM19}.

Two notable series of research on musical instrument recognition are those by Gururani et al. \cite{gururani2017mixing, gururani2018instrument, DBLP:conf/ismir/GururaniSL19} and Hung et al. \cite{yun_ning_hung_2018_1492363, hung2018learning, hung_icassp}. In \cite{DBLP:conf/ismir/GururaniSL19}, the authors found that the use of temporal attention can improve the performance of instrument recognition compared to that of temporal max-pooling adopted in \cite{gururani2018instrument}. On the other hand, the series of work by Hung et al. focused on the representation of pitch and timbre and its role in improving instrument recognition. Their most recent work \cite{hung_icassp} found that the use of joint representation of pitch and timbre outperforms the use of pitch conditioning only \cite{yun_ning_hung_2018_1492363}, though the latter itself still improves the classification performance compared to when no pitch conditioning is used. 

\subsection{Music Audio Tagging}
Music audio tagging is a subfield of MIR closely related to musical instrument recognition. While musical instrument recognition focuses specifically on identifying the presence or absence of the instruments, music audio tagging typically extends their interests to that of moods, tones, genre, etc. Recent works in music audio tagging have also been dominated by deep learning approaches, particularly that of CNN (e.g. \cite{keunwoo_choi_2016_1416254, romain_hennequin_2018_1492499, DBLP:conf/ismir/ChoiLPN19, pons2016experimenting, jordi_pons_2018_1492497, pons2019musicnn}). 

Notably, a series of work by Pons et al. \cite{pons2016experimenting, jordi_pons_2018_1492497, pons2019musicnn} advocated for the use of musically-motivated kernels in CNN, as opposed to the usual square kernels, the latter originating from the image processing fields but carrying less musical interpretability. The authors have shown that the use of musically-motivated kernels can achieve comparable performance to that of traditional kernels.

\subsection{Attention Mechanism}
The attention mechanism first appears in literature in the context of machine translation \cite{bahdanau2014neural, luong2015effective}, particularly to aid in sequence-to-sequence modeling. The attention model used by Bahdanau et al. \cite{bahdanau2014neural} and Luong et al. \cite{luong2015effective} revolves around giving some form of `score' to a particular instance of the input. In \cite{xu2015show}, this model of attention is called the `deterministic soft attention' as the entire input is attended to by the model before the `score' for each `location' is determined. Since then, the use of soft attention has seen some success in multi-instance learning, e.g. \cite{xu2018large, kong2018audio, yu2018multi}, where attention is used in a many-to-one context to perform weighted aggregation in audio set classification. The musical instrument classification work by Gururani et al. \cite{DBLP:conf/ismir/GururaniSL19} also uses this form of attention.

Another model of attention only allows the network to focus on some `location' on the input at any point in time, before stochastically attending the next location. This is the family of attention model adopted by \cite{mnih2014recurrent} and the `stochastic hard attention' in \cite{xu2015show}. To the best of our knowledge, this type of attention has not been adopted in any MIR system.

\section{Method} \label{method}

Let $\mathbf{X}_i\in\mathbb{R}^{T \times F}$ be a log-scaled mel-spectrogram representation of a recording $i$ and $\bm{y}_i\in\mathbb{R}^{C}$ be the corresponding label for each of the $C=20$ instrument classes. In this project, we considered two models: AttentionMIC and Sightreader. AttentionMIC is a sliding-window CNN model heavily based on \cite{DBLP:conf/ismir/GururaniSL19} and uses deterministic attention. Sightreader is adapted from the model in \cite{mnih2014recurrent} and uses stochastic attention. We use $T = 998$ frames and $F = 64$ mel-frequency bins for both models. The details of each model are as follows.

\subsection{AttentionMIC Model Architecture}

\begin{figure}
    \centering
    \includegraphics[width=0.5\textwidth]{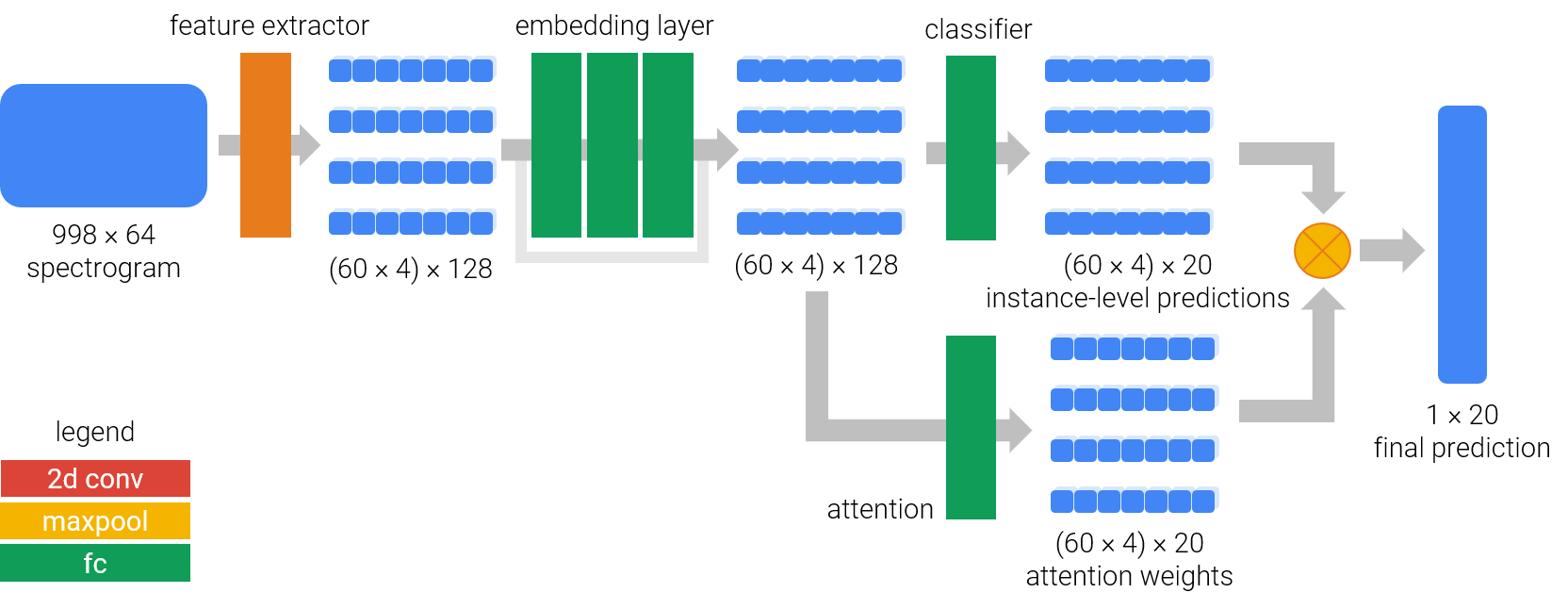}
    \caption{AttentionMIC model architecture}
    \label{fig:AttentionMIC-model}
    \vspace{\baselineskip}
    \centering
    \includegraphics[width=0.5\textwidth]{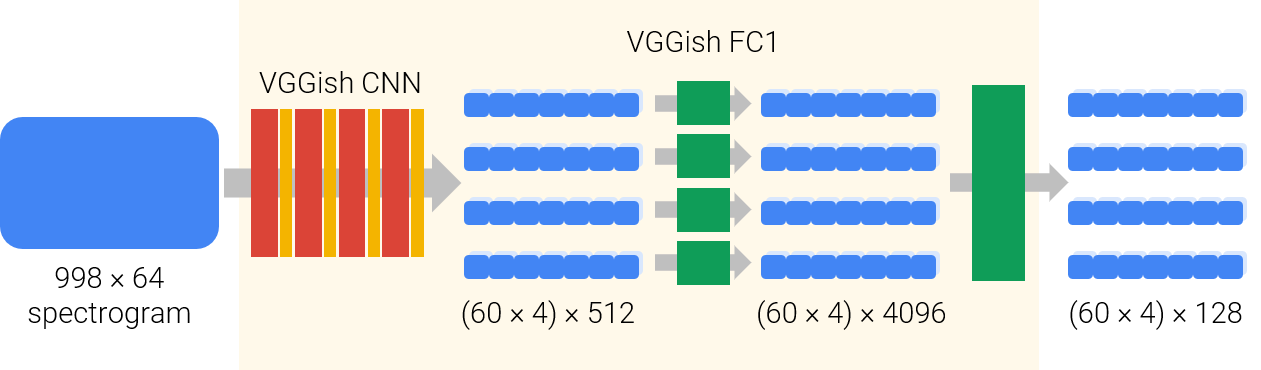}
    \caption{AttentionMIC feature extractor}
    \label{fig:AttentionMIC-feature}
\end{figure}

The AttentionMIC model consists of three main modules: the feature extractor, the embedding layer, and the classifier. Figure \ref{fig:AttentionMIC-model} shows the overall model architecture. 

\subsubsection{Feature Extraction}
The feature extractor is a partially pretrained network based on the VGGish model \cite{VGGish}, as shown in Figure \ref{fig:AttentionMIC-feature}. First, the spectrogram is passed through the VGGish CNN layers to obtain
\begin{equation}
    \mathbf{U}_i = \left[\bm{u}_{i}^1,\bm{u}_{i}^2,\dots,\bm{u}_{i}^{F'}\right]= f_\text{CNN}\left(\mathbf{X}_i\right) \in\mathbb{R}^{\left(T' \times F'\right) \times 512}.
\end{equation}
where $T' = 60$, $F' = 4$, and $\bm{u}_{i}^f\in\mathbb{R}^{\left(T'\times 1\right)\times512}$. The original VGGish fully-connected (FC) layers however requires flattening of the feature vectors across timbral-temporal instances, rendering it impossible to distinguish the contribution from each timbral-temporal instance. As such, we only used a modified version of the first FC layer, by passing each $\bm{u}_{i}^f$ through the segment of the first FC layer in the original VGGish that is responsible for that timbral ordinate, then another trainable FC layer to obtain a set of $\left(T'\times F'\right)$ 128-dimensional feature vectors. This allows each feature vector to still correspond to a particular timbral-temporal region in the spectrogram. 

\subsubsection{Embedding Layers}
The embedding layers consist of three FC layer with 128 nodes and a skip connection. The embedding layers take the feature extractor output and produce embedded feature vector at timbral-temporal coordinates $(t,f)$
\begin{equation}
    \bm{v}_{i}^{t,f}=f_\text{emb}(\bm{u}_i^{t,f})\in\mathbb{R}^{128}.
\end{equation}

In a variant of the AttentionMIC model, we also experimented with appending a frequency identifier to the embedded feature vectors. This is done by appending a one-hot vector $\bm{c}_i^{t,f}\in\left\{0,1\right\}^{F'}$ to the embedded feature, where $\left[\bm{c}_i^{t,f}\right]_f = 1$ and zero elsewhere.

\subsubsection{Attention-based Classifier}
To obtain the overall prediction for the recording, we first obtain instance-level prediction from the embedded feature via
\begin{equation}
    \check{\bm{y}}_i^{t,f} = f_\text{inst}\left(\bm{v}_{i}^{t,f}\right) \in [0,1]^{C} 
\end{equation}
and calculate a normalized attention score for each instance
\begin{equation}
    \bm{\alpha}_i^{t,f} = \dfrac{f_\text{att}\left(\bm{v}_{i}^{t,f}\right)}{\sum_{\tau}\sum_{\phi} f_\text{att}\left(\bm{v}_{i}^{\tau,\phi}\right)} \in [0,1]^{C}
\end{equation}
where $f_\text{inst}(\cdot)$ and $f_\text{att}(\cdot)$ are FC layers. 

The weighted average of predictions across timbral-temporal instances
\begin{equation}
    \hat{\bm{y}}_i = \sum_{t}\sum_f \bm{\alpha}_i^{t,f}\check{\bm{y}}_i^{t,f} \in [0,1]^{C}
\end{equation}
thus gives the final prediction for recording $i$. 

\subsection{Sightreader Model Architecture}
\begin{figure}[!t]
    \centering
    \includegraphics[width=0.35\textwidth]{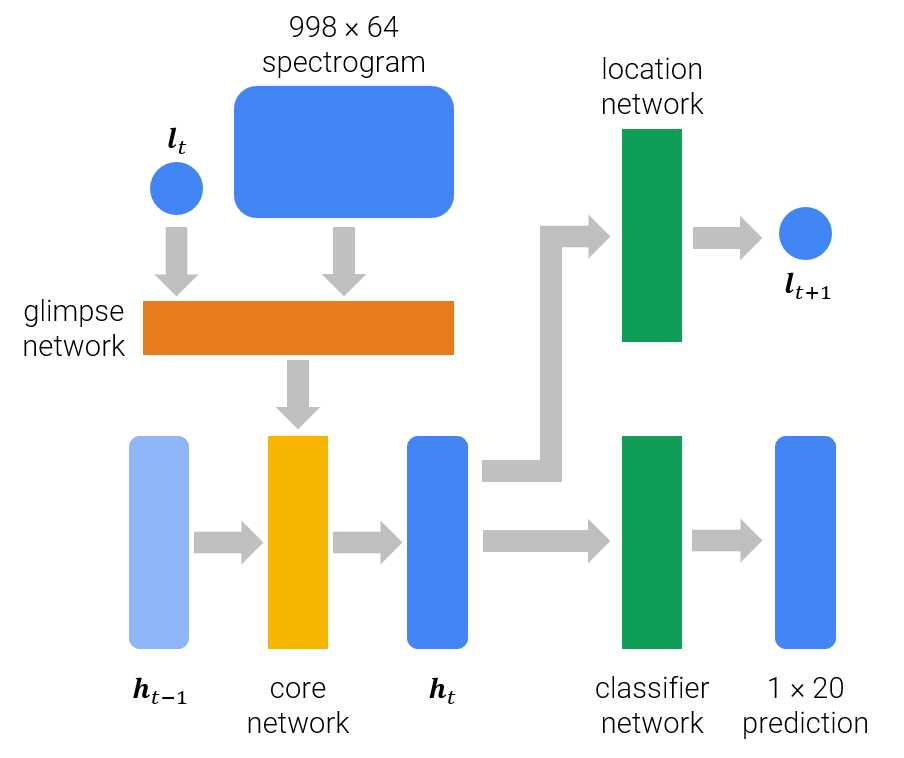}
    \caption{Sightreader model architecture}
    \label{fig:Sightreader-model}
\end{figure}

The Sightreader model is based on the model presented in \cite{mnih2014recurrent}, which uses the recurrent attention model (RAM). Figure \ref{fig:Sightreader-model} shows an overview of the model architecture. Sightreader consists of four main modules: the glimpse network, the core network, the location network, and the classifier network. For simplicity, we drop the recording index $i$ for this section. The details of the Sightreader model are as follows.

\subsubsection{Glimpse network}
The glimpse network takes in the spectrogram $\mathbf{X}$ and some normalized location $\bm{l}_j\in[-1,1]^2$. The glimpse network uses a glimpse sensor to extract $K$ patches, $\bm{x}_{j,k}$, of the spectrogram centered at $\bm{l}_j$, each of sizes $\bm{s}_{k}$. These patches are flattened and passed through a FC layer to obtain
\begin{equation}
    \bm{q}_j = f_\text{sensor}\left(\mathbf{X}, \bm{l}_j; \left\{\bm{s}_k\right\}\right) \in\mathbb{R}^{128}
\end{equation}
which is then combined with a representation from the location vector $f_\text{l}\left(\bm{l}_j\right)\in\mathbb{R}^{128}$ to obtain the feature vector
\begin{equation}
    \bm{\rho}_{j} = f_\text{glimpse}\left(\bm{q}_j + f_\text{l}\left(\bm{l}_j\right)\right)
\end{equation}
which is passed to the core network.

\subsubsection{Core network}
The core network is a gated recurrent unit (GRU) with 256 hidden units. The core network takes in the feature vector $\bm{\rho}_{j}$ and produces a hidden vector $\bm{h}_j\in\mathbb{R}^{256}$ which is passed to both the location network and the classifier network. In the original model by \cite{mnih2014recurrent}, this is a recurrent neural network (RNN) in but we found that GRU tends to perform better (see \ref{eval}), likely due to the memory properties.

\subsubsection{Location network}
The location network takes the hidden vector $\bm{h}_j$ and decide `where' the glimpse network will look at at the next time step. The location network has a FC layer which produces a nominal location $\bm{\mu}=f_\text{loc}(\bm{h}_j)\in[-1,1]^2$. The next location $\bm{l}_{j+1}$ is then found by sampling a from $\mathcal{N}_2\left(\bm{\mu},\sigma^2\mathbf{I}\right)$ with $\sigma=0.17$ arbitrarily chosen.
\subsubsection{Classifier network}
The classifier network takes the hidden vector $\bm{h}_j$ and produces a preliminary prediction $\check{y}_j = f_\text{pred}(\bm{h}_j)\in[0,1]^{C}$ using a FC layer. While only the prediction from the last time step is used as the final prediction for testing, the preliminary predictions are used during training to calculate the reward signal.

\subsection{Loss function}
We experimented with binary cross-entropy (BCE) and focal loss (FL) \cite{lin2017focal} for classification loss. Both BCE and FL are adjusted for partial labels by only using known labels \cite{durand2019learning}, that is, for recording $i$,
\begin{equation}
    \text{CLoss}_i = \dfrac{1}{|\mathcal{C}_i|}\sum_{c \in \mathcal{C}_i} g(\hat{y}_{ic}; y_{ic})
\end{equation}
where $\mathcal{C}_i = \left\{c : \text{label for class $c$ in example $i$ is known}\right\}$, $\hat{y}_{ic}$ and $y_{ic}$ are the prediction and the actual label for class $c$ in example $i$, and $g(\cdot)$ is either BCE or FL. For this project, we use $\gamma = 2$ as the FL modulating factor and the balancing factor is the normalized inverse frequency of positive examples in each instrument class in the provided training set of OpenMIC-2018. 

For AttentionMIC, the model is deterministic; hence the loss function is simply the classification loss. For Sightreader, whose location network is stochastic, however, a hybrid loss function, consisting of the REINFORCE loss and the aforementioned classification loss, has to be used. We refer the reader to \cite{mnih2014recurrent} for a detailed discussion of the hybrid loss function. The cumulative rewards signal used is defined by $R_{ij} = \sum_{k = 1}^{j}r_{ik}$ where $r_{ij}$ is the mean accuracy of the preliminary predictions at time step $j$, adjusted for missing labels. The cumulative reward signal would allow the network to learn a policy for the glimpse trajectory based on the path taken and not simply the location.

Both models are trained with Adam optimization algorithm with a batch size of 32, the learning rate of \num{5e-4} with weight decay, for 250 epochs and early stopping after 10 epochs without improvements in the F1 score on the validation set. We checkpoint the model with the best validation F1 score.

\section{Evaluation} \label{eval}

\subsection{Dataset}
The OpenMIC-2018 dataset is used for all experiments in this project. We used the provided train-test split for reproducibility, and randomly sampled \SI{15}{\percent} of the training set as the validation set. We generated log-scaled mel-frequency spectrograms from the raw audio files provided in the dataset using the same setting as that used to generate the provided features. 

\subsection{AttentionMIC Experiments}

We compared the proposed AttentionMIC model, AttTF, and the frequency-identified variant, AttTFid, to the following models:
\begin{enumerate}
    \item AttT: This model serves as an ablation model by only using temporal attention. The rest of the model is the same as the AttTF models. The model is designed to closely resemble the `ATT' model in \cite{DBLP:conf/ismir/GururaniSL19}.  
    \item FC: A baseline model with a fully-connected layer replacing the attention-based classifier in AttTF.
\end{enumerate}

All models in this experiment are trained using BCE as the loss function.

\subsection{Sightreader Experiments}

We evaluate the following variants of the Sightreader and RAM models \cite{mnih2014recurrent}:

\begin{enumerate}
    \item SR16: Sightreader with 16 glimpses trained using BCE.
    \item SR16-FL: Sightreader with 16 glimpses trained using focal loss.
    \item RAM16-RNN: This model closely resembles the original model from \cite{mnih2014recurrent} with the classifier modified for multi-label classification. The model has 16 glimpses with kernel size 12 by 12 pixels. BCE is used for classification loss.
    \item RAM16-GRU: The same model as RAM-RNN but with the core network changed to GRU. 
\end{enumerate}

\subsection{Metrics}
For all experiments, we report both the macro-averaged F1 score as well as the instrument-wise F1 scores achieved by each model for comparison. The instrument-wise scores are averaged across 5 seeds for clarity.

\section{Results and Discussion} \label{results}

\begin{figure*}[!t]
\centering
\includegraphics[width=0.92\textwidth]{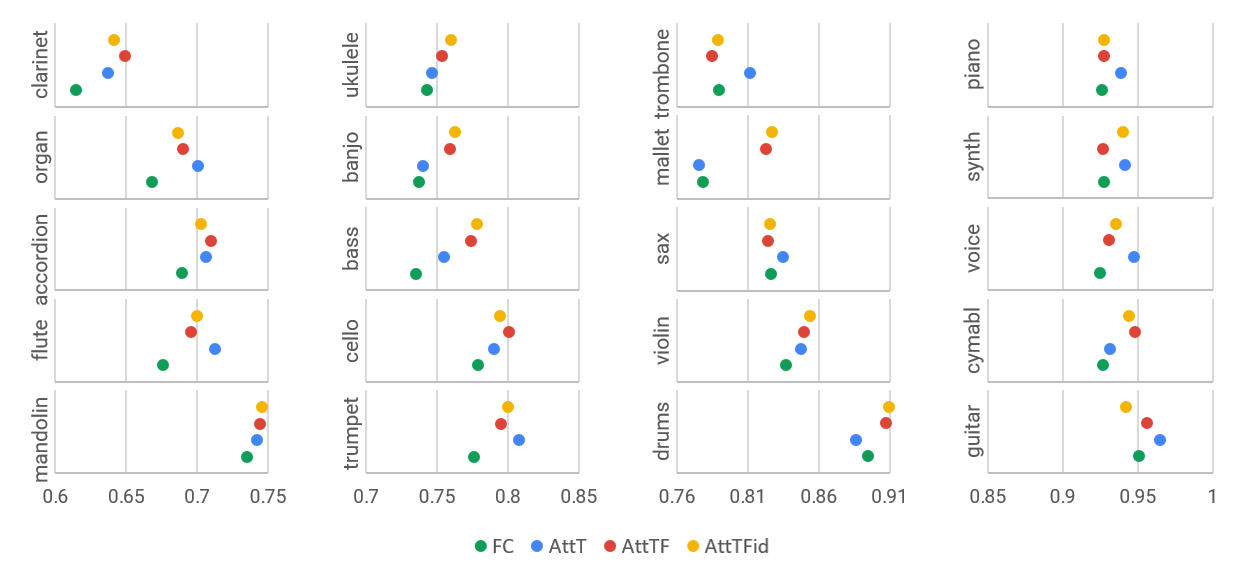}
\caption{Instrument-wise F1 scores for models evaluated in the AttentionMIC experiments}
\label{fig:AttentionMIC-inst}

\includegraphics[width=0.92\textwidth]{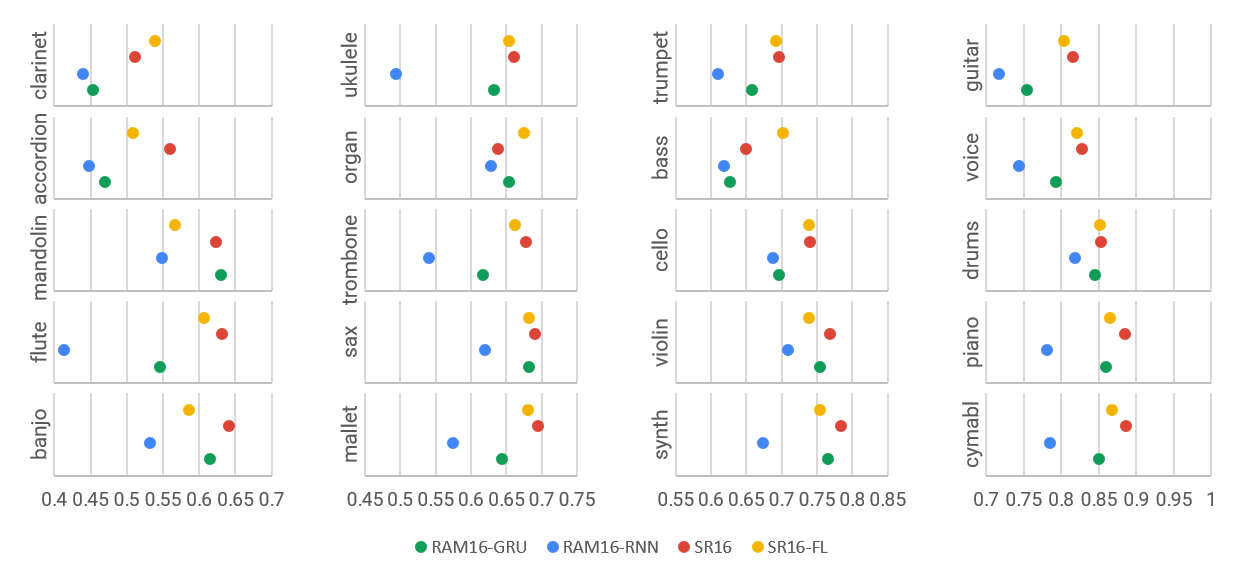}
\caption{Instrument-wise F1 scores for models evaluated in the Sightreader experiments}
\label{fig:Sightreader-inst}
\end{figure*}

The overall and instrument-wise results for the AttentionMIC experiments are shown in Figures \ref{fig:AttentionMIC-F1} and \ref{fig:AttentionMIC-inst}, respectively. Similarly, the overall and instrument-wise results for the Sightreader experiments are shown in Figure \ref{fig:Sightreader-F1} and \ref{fig:Sightreader-inst}.

\subsection{AttentionMIC experiments}

\begin{figure}[!t]
\centering
\includegraphics[height=1.5in]{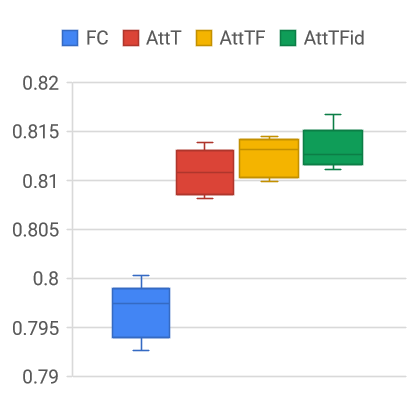}
\caption{Overall F1 scores for models evaluated in the AttentionMIC experiments}
\label{fig:AttentionMIC-F1}
\end{figure}

For the AttentionMIC models, it is clear from Figure \ref{fig:AttentionMIC-F1} that the use of attention results in an improvement in the classification performance compared to the FC model. Overall, the visual attention models (AttTF and AttTFid) performs slightly better than the temporal attention model (AttT), though the improvement is slight. We hypothesize that the improvement is due to AttTF and AttTFid's ability to focus on specific frequency bands where an instrument is active. We further hypothesize that the frequency identifier in AttTFid allows the model to treat instances from different frequency regions differently, thus a slight improvement from the AttTF model. 

It can be seen in Figure \ref{fig:AttentionMIC-inst} that the visual attention models perform better than the temporal attention models in many instrument classes. Interestingly, some instrument classes where the temporal attention model outperforms visual attention models (piano, guitar, synthesizer, organ) are chordal instruments whose range can span several octaves -- hence several frequency instances in the visual attention models. This may perhaps be explained by a major downside of our visual attention models; by treating different frequency bands as independent instances, features that require a large span of frequency could not be effectively extracted -- an ability that the temporal model is otherwise capable of. 

Furthermore, we also observe that within an instrument class, all models tend to have similar performances. This could suggest that some instruments are inherently harder to classify than others. This may be due to the inherent nature of its tone and its musical role in a composition, and/or the positive-negative imbalance in the OpenMIC-2018 dataset.

\subsection{Sightreader experiments}

\begin{figure}[!t]
\centering
\includegraphics[height=1.5in]{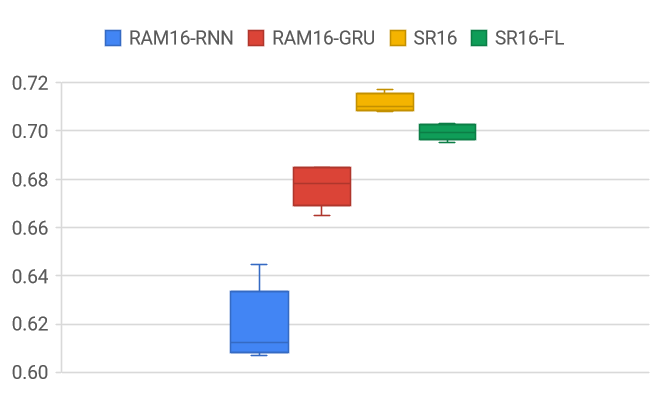}
\caption{Overall F1 scores for models evaluated in the Sightreader experiments}
\label{fig:Sightreader-F1}
\end{figure}

From Figure \ref{fig:Sightreader-F1}, the use of GRU in lieu of RNN as the core network in the RAM model clearly results in a significant improvement in the F1 score. We hypothesize that this is due to GRU's ability to `memorize' its past state unlike RNN. This also justifies the use of GRU for the Sightreader model instead of RNN.

Between the RAM16-GRU model and the SR16 model, we see another significant improvement. While the RAM16-GRU model uses a square kernel, the SR16 model uses a set of musically motivated kernels based on the \textit{musicnn} network \cite{pons2019musicnn}. We hypothesize that such kernels allow the SR16 model to capture more musically meaningful features, thus the superior performance.

In the SR16-FL model, we experimented with the focal loss as a classification loss. However, we see a drop in performance across most instrument classes save for clarinet, organ, and bass. Since focal loss is designed to give larger gradient to `harder' examples, we hypothesize that the use of focal loss may sacrifice the overall quality of the embedding space in favor of these `hard' examples, hence the drop in performance. 

Overall, however, the Sightreader models consistently perform worse than any AttentionMIC model by roughly \SI{10}{\percent}. While the reason behind the inferior performance is unclear, the different feature extraction procedures in the two models could offer some explanation. The AttentionMIC model extracts features from the entire spectrogram first, then choose which instance to prioritize. The Sightreader model, on the other hand, chooses where to look and only extract information around those regions. As a result, some regions of the spectrogram may be ignored entirely by the Sightreader model -- thus a potential loss of input information which could otherwise be useful.

\section{Conclusion} \label{conclusion}

In conclusion, we explored two approaches of visual attention to improve musical instrument classification. First, using deterministic attention in the AttentionMIC model, we find a slight improvement in the classification performance from the temporal attention model to the visual attention models. In the second approach, using stochastic attention in the Sightreader model, we see the importance of musically motivated kernels, which allows a more effective feature extraction compared to the square kernels conventionally used in image processing. 

Moving forwards, the sliding-window approach may still be more appropriate for the multi-instrument music where the salient features can be hard to locate. While we are currently limited by the available dataset, we believe that with a sufficiently large and well-annotated dataset, modifying the AttentionMIC model with the musically motivated kernel could prove beneficial in the future.

\bibliographystyle{IEEEtran}
\bibliography{bib}

\end{document}